\documentclass{article}
\usepackage{amsmath}

\usepackage[preprint]{neurips_2026}
\usepackage{natbib}
\usepackage[utf8]{inputenc} 
\usepackage[T1]{fontenc}    
\usepackage{hyperref}       
\usepackage{url}            
\usepackage{booktabs}       
\usepackage{amsfonts}       
\usepackage{nicefrac}       
\usepackage{graphicx}
\usepackage{float}
\usepackage{microtype}      
\usepackage{xcolor}         
\usepackage{multirow}
\usepackage[most]{tcolorbox}
\usepackage{pgfplots}
\pgfplotsset{compat=1.18}
\usepackage{xcolor}
\usepgfplotslibrary{groupplots}

\newtcolorbox[auto counter]{takeaway}{
  rounded corners=south,
  arc=2pt,                          
  colback=gray!5,                   
  colframe=gray!25,                 
  boxrule=0.5pt,                    
  left=10pt, right=10pt, top=8pt, bottom=8pt, 
  before upper={\textbf{Property \thetcbcounter. }} 
}
\title{Lights, Camera, Carbon: Architectural Scaling Laws for Video Generation Energy Consumption}

\author{%
  Nidhal Jegham \\
  University of Rhode Island, Sustainable AI Group \\
  Rhode Island, United States \\
  \texttt{nidhal@sustainableaigroup.com} \\
  \AND
  Boris Gamazaychikov \\
  Sustainable AI Group \\
  Paris, France \\
  \And
  Sasha Luccioni \\
  Sustainable AI Group \\
  Montreal, Canada\\
}

\begin{document}

\maketitle
\vspace{-1em}

\begin{figure}[h!]
    \vspace{-1em}

    \centering
    \includegraphics[width=1\linewidth]{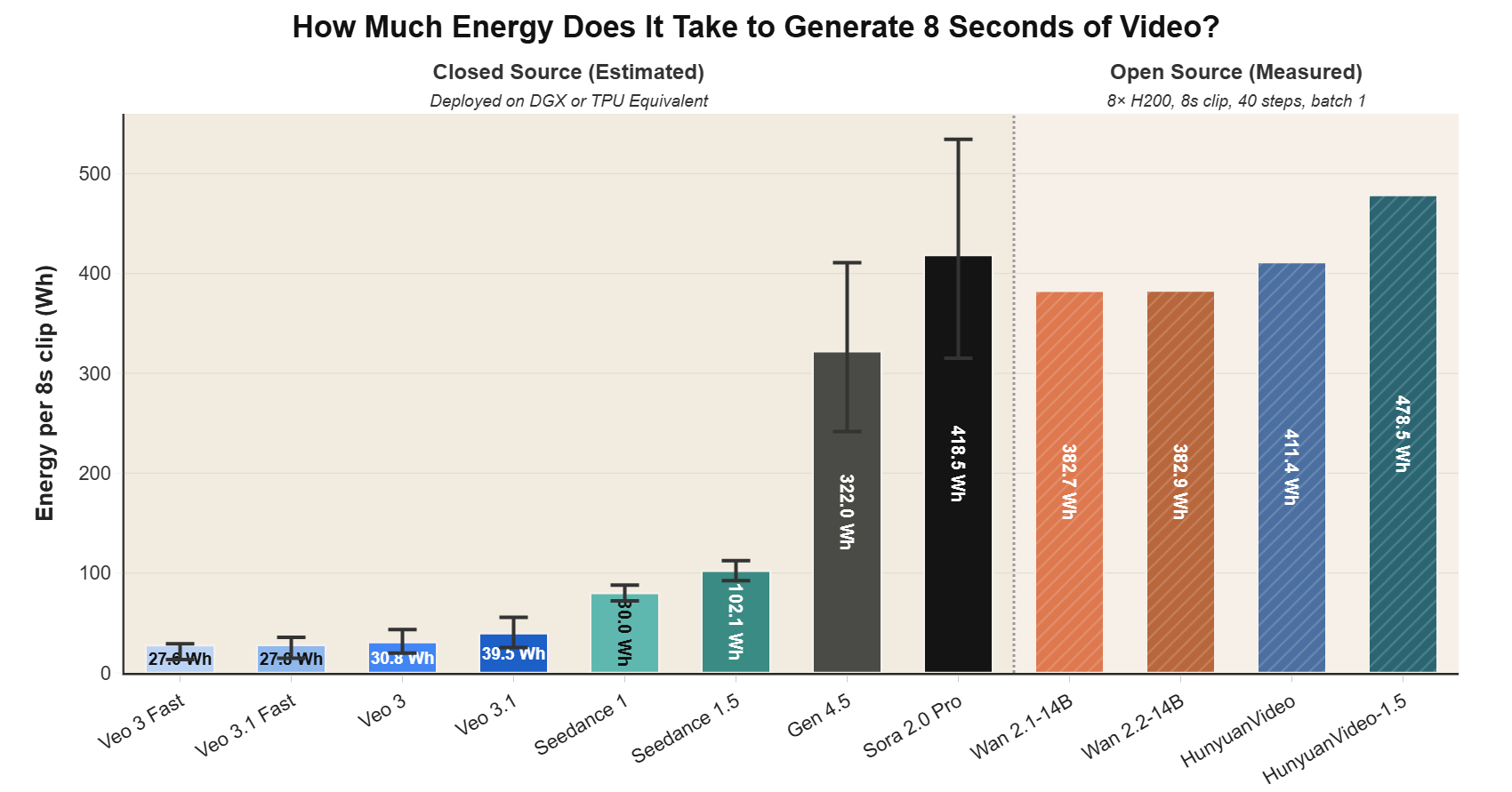}
    \vspace{-1em}

    \caption{Measured open-model energy consumption versus estimated proprietary-model energy consumption for an 8-second 720p generation.}
    \label{fig:abs}
\end{figure}
\vspace{-1em}

\begin{abstract}
  We present a bidirectional framework for estimating the energy consumption of text-to-video (T2V) and text-to-video-audio (T2VA) models from architectural first principles and observable generation parameters such as resolution and duration, requiring no access to weights, model size, or implementation details. \emph{Forward}, it predicts energy from generation parameters and architectural principles; \emph{backward}, it recovers architectural scaling behavior from observed inference times, with accuracy serving as a criterion for architectural validity. Building on the established compute-bound nature of video diffusion models, we demonstrate that each model's energy profile obeys theoretically derived scaling laws, decomposing into quadratic and linear terms whose coefficients directly reflect the underlying architectural complexity. 
  Validated across six open-source models spanning 8.3B–27B parameters and three GPU configurations, this decomposition achieves below 3\% MAPE across all architectures. This approach offers a standardized, empirically and theoretically grounded framework for sustainability benchmarking across T2V models and architectures.
\end{abstract}


\section{Introduction} \label{sec_intro}

Recent years have seen a rise in the widescale deployment of machine learning (ML) systems in a variety of user-facing applications and tasks, generating responses to user queries in modalities ranging from text~\cite{openai2024gpt4technicalreport} to audio~\cite{kim2021conditional} and images~\cite{rombach2022high}. However, the  rapidly growing environmental impacts associated with this growth are an increasingly important topic to examine and factor into the development and deployment of ML systems~\cite{cowls2023ai}. As such, the sustainability and, more specifically, the energy demands of different ML systems have become a nascent but rapidly developing field of study. Starting with the seminal work of Strubell et al in 2019~\cite{strubell2019energy}, the subsequent years of scholarship have shed more light on different ML tasks, the relative contribution of the different stages of the ML life cycle, as well as the factors that influence them, mainly focusing on text-based models~\cite{luccioni2022estimating, fernandez2025energy,fernandezevaluating, morrison2025holistically,pronk2025benchmarking,jegham2025hungry}. Most recently, the topic has been broadened to also include other modalities such as image generation and speech-to-text~\cite{Luccioni_2024, ulan2025talk}.

Given that text-to-video (T2V) generation is a relatively novel ML task, its energy consumption has yet to be analyzed in depth in an empirical way. The few existing studies that examine this modality have focused on specific models~\cite{li2024carbon} or a few open model families~\cite{delavande2025video} -- these studies have found that video generation is not simply a scaled-up version of image generation, mainly because it requires \textit{iterative denoising} across both spatial and temporal dimensions, which generate hundreds of frames per output (as opposed to a single image, in the case of image generation). The most complete study to date, by Delavande et al.~\cite{delavande2025video}, found that AI video generation operates in a \emph{compute-bound regime}, with its energy requirements and latency scaling near-quadratically as the resolution and length of output videos increases. Comparing video generation with other modalities, they found that generating a single short video can consume approximately 90 $Wh$ of energy, making it 30 times more costly than image generation and over 2,000 times more costly than text generation.

In the current study, we go beyond previous analyses to theoretically derive and empirically validate the scaling laws governing energy consumption of video generation models from architectural first principles across a broader set of architectures, hardware configurations, and experimental conditions. Building on the established compute-bound nature of video diffusion models, we develop a bidirectional framework that predicts energy from observable generation parameters -- such as resolution, video duration, and batch size -- and recovers architectural scaling behavior from inference times, requiring no access to weights, model size, or implementation details. We additionally characterize the role of architectural decisions over parameter count and model size, the scaling laws of audio generation overhead, and the energy implications of multi-GPU inference. As a case study demonstrating the framework's broader applicability, we apply it to estimate energy consumption for eight proprietary T2V systems for which direct measurement is infeasible.

In order to present these findings, we first start with background information regarding video generation (Sec.~\ref{sec_preliminaries}), then present both the theoretical foundations and empirical methodology of our scientific approach (Sec.~\ref{sec_methodology}). We then present the results of different factors that we tested in our experiments and their influence on GPU energy consumption in Sec.~\ref{sec_results}. Finally, we discuss the significance of our findings and propose future work in Sec~\ref{sec_discussion}.

\section{Background}
\label{sec_preliminaries}

Video generation diffusion models are characterized by three main computational components: a Variational Autoencoder (VAE) encoder, a denoising network, and a VAE decoder. The encoder maps the input into a compressed latent space; the denoising network iteratively refines the latent representation over a specified number of steps; and the decoder maps the latent back to pixel space.

\paragraph{Notations}  Let \textit{W} and \textit{H} denote the width and height of the generation video in pixels and \textit{F} the number of frames defined as $F=L \times FPS + 1$, where \textit{L} is the duration of the video in seconds and FPS refers to Frames Per Second. \textit{B} denotes batch size (or how many videos are being generated at the same time), \textit{$L_{text}$} denotes the input prompt length in tokens, and \textit{S} denotes the number of denoising steps during the denoising stage. We define the spatiotemporal volume, or the total number of pixels in the video as $T =  \frac{(H \times W \times F)}{1000}$ (presented over 1000 for numerical stability).

\subsection{Architectural Decomposition}

The denoising network constitutes the dominant computational cost of the T2V pipeline. At each of the $S$ steps, it processes a compressed latent tensor derived from $T$ pixels through a sequence of operations repeated across transformer blocks. Self-attention, in which every token attends to every other token across space and time, scales as $\mathcal{O}(T^2)$. Cross-attention, where each spatio-temporal token attends to text tokens to condition the generation on the input prompt, scales as $\mathcal{O}(T \cdot L_{text})$ with $L_{text}$ being the input length in tokens . The feed-forward network (FFN), which applies linear projections and non-linear activations independently to each token, also scales as $\mathcal{O}(T)$. \footnote{Some architectures implement full spatio-temporal self-attention, while others decouple it into spatial self-attention, scaling as $\mathcal{O}(F \cdot (HW)^2)$, and temporal self-attention, scaling as $\mathcal{O}(HW \cdot F^2)$, to reduce the $\mathcal{O}(T^2)$ cost.} 

Additionally, models that support native audio generation either concatenate audio and video tokens into a joint sequence of size $T + T_{audio}$, processed through shared self-attention and scaling as $\mathcal{O}((T + T_{audio})^2)$, or maintain separate streams that interact via cross-attention between modalities, scaling as $\mathcal{O}(T \cdot T_{audio})$, in addition to cross-attention with the input prompt. 
Finally, the VAE decoder reconstructs the full-resolution video from the final latent, mapping it back to pixel space. Both encoder and decoder rely primarily on 3D causal convolutions, which scale as $\mathcal{O}(T)$, where each frame can only attend to previous frames, ensuring temporal order. \footnote{Attention can additionally be employed during the decoding phase for video consistency, scaling as $\mathcal{O}(T^2)$, as in the case of the VAE decoder of HunyuanVideo \cite{kong2024hunyuanvideo}}

\subsection{Compute-bound Nature of Diffusion Models}

A key property of GPU workloads is that they can be either \emph{compute-bound}, \emph{memory-bound}, both, or neither in poorly optimized kernels. In memory-bound workloads, power draw fluctuates with memory access patterns, decoupling inference time from computational complexity. In compute-bound workloads, arithmetic units are saturated, driving power to near the hardware's theoretical maximum -- referred to as Thermal Design Power (TDP) -- and maintaining it approximately constant throughout inference, such that Energy $(E) = P \times t$ where $P \approx 80-100\%$ of the TDP.  Unlike LLMs, which are memory-bound  due to the sequential nature of autoregressive decoding~\citep{recasens2025mind}, diffusion models have previously been shown to be predominantly compute-bound, as the denoising network processes the full latent tensor in parallel at each step~\cite{delavande2025video}. However, the VAE decoding phase in video generation can also be memory-bound~\cite{wang2026eliminating} due to the large intermediate feature maps produced when decoding high-resolution, long-duration latent tensors. However, denoising is still the primary driver of energy consumption; thus, video generation is predominantly compute-bound \cite{delavande2025video}.

\section{Scientific Approach}
\label{sec_methodology}
The objective of our study is to examine and formalize the energy consumption scaling laws for AI video generation across different models and configurations. By deriving scaling laws from architectural first principles, we develop a framework that operates bidirectionally: \emph{forward}, deriving energy consumption from architectural first principles and observable generation parameters; and \emph{backward}, recovering architectural scaling behavior from observed inference time. This correct recovery is only possible when the assumed scaling law reflects the true architectural complexity, making fit quality itself a validation of the architectural assumptions. As such, Section~\ref{subsec_theoretical} derives the general energy formula from architectural complexity analysis, establishing the functional form and physical interpretation of each term. Consequently, Section~\ref{subsec_experimental} describes the measurement setup, experimental design, and fitting procedure used to identify model-specific coefficients and validate the framework across open models. Crucially, this backward capability also enables energy estimation for black-box, proprietary video generation models employed by the majority of users.

\subsection{Theoretical Foundation} \label{subsec_theoretical}
Section~\ref{sec_preliminaries} established that diffusion models are predominantly compute-bound, with arithmetic units saturated throughout inference. Two consequences follow. First, power draw remains approximately at TDP, independent of parameter count or model size. Second, under fixed hardware, inference time is proportional to the number of floating-point operations performed, which is itself a deterministic function of generation parameters such as resolution, frame count, and number of denoising steps. Together, these imply that energy consumption is a deterministic function of the same parameters (up to the constant p), and that any predictor of generation time is equivalently a predictor of energy.\footnote{The VAE decoding phase may be memory-bound, introducing a lower average power draw $P_v$ that partially contaminates the first observation. However, the denoising phase dominates total generation time and energy~\cite{delavande2025video}, so these deviations are second-order effects and energy consumption remains a predominantly deterministic function of generation parameters.} We accordingly decompose total energy per generation as $E = E_{\text{denoise}} + E_{\text{VAE}}$.

At each of the $S$ denoising steps, the network processes $T$ tokens through self-attention scaling as $\mathcal{O}(T^2)$, cross-attention to text conditioning scaling as $\mathcal{O}(T \cdot L_{text})$, and the FFN scaling as $\mathcal{O}(T)$. Since input prompts are typically padded to maximum length \cite{kong2024hunyuanvideo,ltx2}, $L_{text}$ is a constant and therefore absorbed by the FFN term. For models with joint audio generation, three additional terms arise: audio self-attention scaling as $\mathcal{O}(F^2)$, audio-visual cross-attention scaling as $\mathcal{O}(T \cdot F)$, and audio FFN scaling as $\mathcal{O}(F)$ with the text input cross-attention absorbed by the FFN. Since videos in a batch are generated independently without attending to one another, denoising steps are sequential and independent, and the total denoising energy scales linearly with both $B$ and $S$, giving:
\begin{small}
\begin{center}
\begin{multline*}
E_{\text{denoise}} = S \cdot B \cdot \Big(
\underbrace{N_v \cdot T^2}_{\text{video self-attn}}+
\underbrace{M_v \cdot T}_{\substack{\text{video FFN}}}+
\underbrace{N_a \cdot F^2}_{\text{audio self-attn}}+
\underbrace{M_a \cdot F}_{\substack{\text{audio FFN}}}+
\underbrace{C_{av} \cdot T \cdot F}_{\text{audio-video cross-attn}}
\Big) + \underbrace{G \cdot B}_{\text{overhead}}
\end{multline*}
\end{center}
\end{small}

where $N_v$ captures the quadratic video self-attention cost, $M_v$ the linear FFN cost over video tokens and text cross-attention,
$N_a$ the quadratic audio self-attention cost, $M_a$ the linear FFN cost over audio tokens and text cross-attention, $C_{av}$ the audio-video cross-attention cost, and $G$ a fixed overhead.

The VAE relies on causal convolutions, where each output depends only on current and previous frames, scaling as $\mathcal{O}(T)$. It can employ attention mechanisms that scale as $\mathcal{O}(T^2)$ for temporal consistency, as discussed previously. As with the denoising phase, batch independence yields linear scaling with $B$, giving:
$E_{\text{VAE}}$ = $B \cdot \Big(
\underbrace{N \cdot T^2}_{\text{self-attn}} + 
\underbrace{M \cdot T}_{\text{convolutions}} + 
\underbrace{G}_{\text{overhead}}
\Big)$

where $N$ captures the attention cost, $M$ the 3D convolution networks cost, and $G$ the fixed overhead. In order to empirically validate these theoretical derivations, we measure the energy consumption of multiple models on different generation parameters and fit these scaling laws' terms to the energy consumption profile of these models.

\subsection{Experimental Methodology} \label{subsec_experimental}

In order to measure energy consumption during the process of video generation, we run a series of experiments
while using the \texttt{pyNVML} package to gather real-time energy measurements~\footnote{We only account for GPU energy consumption in our experiments, since this accounts for the majority of the energy footprint in predominantly compute-bound processes.}. To ensure consistent testing, we used a fixed list of prompts from ArtificialAnalysis across all models~\cite{ArtificialAnalysis2026} -- the full prompt list can be found in Appendix~\ref{appendix_prompts}. During each video generation, we log the power draw, GPU utilization, and memory utilization every 100 milliseconds. We conduct the experiments on 3 different GPU configurations in order to test the impact of hardware choice on energy consumption: a \emph{DGX H200} equipped with 8 Nvidia \emph{H200 SXM} GPUs, denoted as $8H2$ in Table~\ref{table:list_models}; a single \emph{H200 SXM} ($H2$); and a single \emph{B200} ($B2$). 
During experimentation, we separate generation phases such as denoising and VAE decoding to ensure phase-level analysis, and temporal tiling was enabled for HunyuanVideo, HunyuanVideo 1.5, and LTX-2, where VAE decoding exceeded GPU memory capacity at the tested resolutions. Each generation is preceded by two warm up runs to stabilize operating conditions, with a 5-second idle period between runs to dissipate residual thermal load.

Rather than exhaustively testing every combination across all models, we designed the experiments to target specific terms in the derived formulas (e.g., varying number of frames while holding other parameters constant) described in Section~\ref{subsec_theoretical}. While all models are tested across varying frame counts and denoising steps as the primary isolation dimensions, \emph{resolution} is varied for HunyuanVideo and LTX-2 Pro to study the robustness of the scaling laws across different spatial configurations. \emph{Batch size} is varied for HunyuanVideo to study its linearity; \emph{GPU type} and \emph{count} are varied for HunyuanVideo 1.5, Wan 2.1, Wan 2.2, and LTX-2 to characterize hardware scaling. Finally, LTX-2 is tested \emph{with and without audio} by decoupling the audio module to isolate its energy cost. The full list and characteristics of the models that were tested are presented in Table~\ref{table:list_models}.

\begin{table}[h!]
\centering
\begin{tabular}{lllllll}
\textbf{Model}   & \textbf{Size} & \textbf{Modalities} & \textbf{Count} & \textbf{GPU types} & \textbf{FPS} \\ \hline
HunyuanVideo~\cite{HY}   & 13B  & T2V & 67  & $8H2$ & 24 \\ \hline    
HunyuanVideo-1.5~\cite{HY1.5}  & 8.3B  & \begin{tabular}[c]{@{}l@{}}T2V \end{tabular} & 15   & $B2$, $H2$, $8H2$ & 24 \\ \hline
LTX-2~\cite{LTXHF}         & 19B                 & \begin{tabular}[c]{@{}l@{}}T2VA T2V\end{tabular} & 80                         & $B2$, $H2$   & 24                    \\ \hline
Wan 2.2~\cite{Wan2.2}                 & 27B                 & \begin{tabular}[c]{@{}l@{}}T2V \end{tabular}                    & 23                         & $B2$, $H2$, $8H2$      & 16                       \\ \hline
Wan 2.1~\cite{Wan2.1}                & 14B                 & \begin{tabular}[c]{@{}l@{}}T2V \end{tabular}                    & 18                         & $B2$, $H2$, $8H2$  & 16
                       
\end{tabular}
\begin{flushleft}

\end{flushleft}
\caption{List of models tested}
\label{table:list_models}
\end{table}
Energy consumption across our tested configurations spans nearly three orders of magnitude, from 2.95 $Wh$ for a 3-second LTX-2-T2V generation at 1024p on $B2$ to 3,364.6 $Wh$ for a 23-second Wan 2.2 generation at 720p with 50 steps on $8H2$. LTX-2 is the most energy-efficient model tested;
Wan variants were the least. At a fixed workload of a 5-second output with 40 steps, per-generation cost varies by an order of magnitude across models. LTX-2-T2V at 1024p consumes 9.1 $Wh$ on $B2$, while at 720p HunyuanVideo-1.5 consumes 57.5 $Wh$ and Wan 2.1/2.2 consume 113.9/114.8 $Wh$ -- equivalent to running a 1,200W air fryer for 27 secs, 2.9 mins, and 5.7 mins~\cite{equivalentsite}. 

The per-generation footprint is striking even relative to text generation: an 8-second 720p Wan 2.2 generation at 40 steps on $H2$ (390 $Wh$) consumes 1,625$\times$ more energy than a Google Gemini text prompt (0.24 $Wh$)~\cite{elsworth2025measuring}, and an 8-second HunyuanVideo-1.5 generation on $8H2$ (478.5 $Wh$) nearly 1,990$\times$ more. Scaling to 100 million such generations (a number reached by Google in 2025~\cite{CNETFlow2024}) corresponds to 39 $GWh$ for Wan 2.2 and 47.9 $GWh$ for HunyuanVideo-1.5 -- the annual electricity consumption of 3,645 and 4,477 average US households~\cite{eia_faq_2024}. Our experiments required 17 GPU hours of total inference, requiring 61.9 $kWh$ while emitting approximately 1.83 $kgCO_2e$ 
~\cite{ElectricityMaps2024}), equivalent to the use of a microwave for 51 hours~\cite{equivalentsite} in terms of energy consumption or the carbon emissions of a 7.25 km drive in a gasoline car~\cite{EPA_GHG_Vehicle_2024}. 

\section{Results}
\label{sec_results}

Having derived the functional form of the energy scaling laws from architectural first principles, our experimental methodology serves to systematically validate and extend this theoretical approach across a comprehensive set of experiments, presented in Table~\ref{table:list_models}. Our results analysis pursues two goals: validating the compute-bound nature of video diffusion models across diverse architectures and generation pipelines -- spanning different model sizes, batch sizes, and GPU configurations -- and confirming that the scaling laws derived from the architectural principles in Section~\ref{subsec_theoretical} hold empirically. We also disentangle the quadratic and linear decomposition of the models tested, demonstrating the bidirectionality of the framework: \emph{forward}, by showing that architecture-derived scaling laws predict energy consumption for open models (in Sec.~\ref{forward}); and \emph{backward}, by showing that this accuracy is only achieved when the scaling law correctly reflects the underlying architecture (in Sec.~\ref{backward}). Finally, by leveraging the derived properties of video generation models from our analysis, we estimate the energy consumption of proprietary models like Sora and Veo (in Sec.~\ref{subsec_prop}). Our experimental design is therefore directly motivated by the formula structure; each configuration (resolution, frame count, batch size, denoising steps) is leveraged specifically to isolate and identify individual terms in the formula, rather than to exhaustively benchmark model performance.

\subsection{The Compute-Bound Nature of Diffusion Models} 
\label{subsec_compute_bound}

Our experiments confirm the compute-bound nature of video generation models: all tested models (Table~\ref{table:list_models}), ranging from 8.3 to 27B parameters, sustain near-TDP power draw across the full generation pipeline, with total average power reaching 80.9--98.7\% of TDP depending on model and hardware. LTX-2 variants warrant separate treatment. They follow a two-stage pipeline: a primary denoising stage at half the target resolution, followed by three denoising steps at full resolution, which reduces token count and improves efficiency at the cost of temporal consistency and fine detail relative to models that denoise at full resolution. However, because the second denoising stage is fused with VAE decoding, the two cannot be separated for phase-level analysis. The overall pipeline nonetheless reaches 80.9--81.1\% $\pm$ 9--12\% of TDP on $B2$ and 86.2--92.1\% $\pm$ 2--3\% on $H2$, with a lower percentage for reduced resolutions and frame counts.

Breaking down energy use by phase, we find the denoising stage is particularly compute-bound: non-LTX models sustained 98.8--98.9\% of TDP on $B2$ GPUs, 98.5--98.6\% on $H2$ GPUs, and 97.8--98.1\% on the $8H2$ station. In fact, significant deviations from TDP were confined to the VAE decoding for models where temporal tiling was enabled due to memory constraints (e.g., HunyuanVideo, HunyuanVideo-1.5, LTX-2), dropping VAE-phase power to as low as 71.6\% of TDP for HunyuanVideo with low resolution and reduced FPS and 83.5--89.5\% for HunyuanVideo-1.5, depending on hardware type. However, for Wan 2.1 and Wan 2.2, where tiling was not required due to architectural efficiencies, VAE decoding reaches 90.4--96.1\% of TDP and contributes only 0.83--3.43\% of total generation energy, rendering its overall contribution to energy usage negligible. For HunyuanVideo, the VAE fraction was more pronounced, ranging from 4.2--25.8\% of total energy, with the upper end occurring at low frame counts and resolutions (480$H$, 640$W$, 30$S$, 121$F$) where the VAE cost is least amortized by denoising. Overall, the denoising phase dominates total time and energy across all models, and the total average power draw per model still runs near or at TDP.

\begin{takeaway}
Video generation is a predominantly compute-bound process, with models sustaining power draw near TDP regardless of model size or architecture. Consequently, energy consumption differences across models are driven predominantly by inference time, and consequently, the number of floating-point operations executed. Therefore, any predictor of inference time is equivalently a predictor of energy consumption.
\end{takeaway}


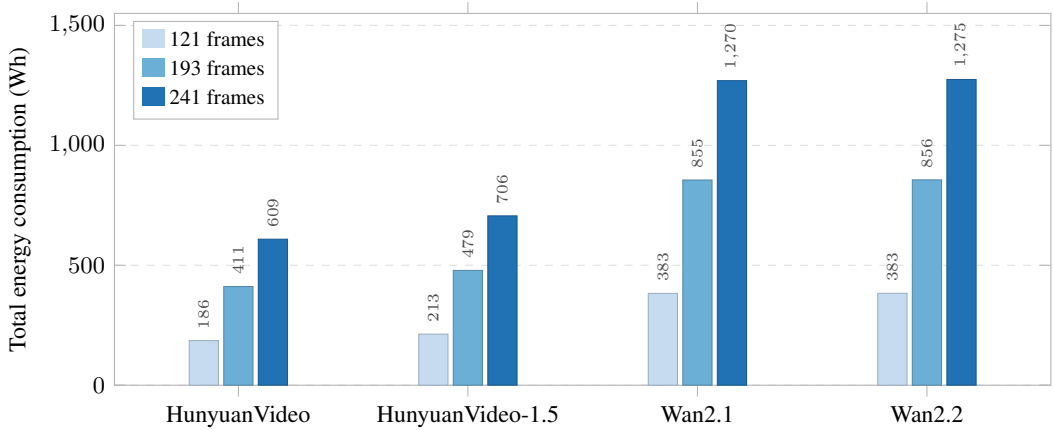
\begin{figure}[h!]
    \centering
    \definecolor{barlow}{HTML}{C6DBEF}
    \definecolor{barmid}{HTML}{6BAED6}
    \definecolor{barhigh}{HTML}{2171B5}
    \begin{tikzpicture}
        \begin{axis}[
            ybar=2pt,
            bar width=11pt,
            width=\linewidth,
            height=6.5cm,
            enlarge x limits=0.18,
            ymin=0,
            ymax=1550,
            ylabel={Total energy consumption (Wh)},
            symbolic x coords={HunyuanVideo, HunyuanVideo-1.5, Wan2.1, Wan2.2},
            xtick=data,
            x tick label style={font=\small},
            ylabel style={font=\small},
            yticklabel style={font=\small},
            legend style={
                at={(0.02,0.98)},
                anchor=north west,
                font=\footnotesize,
                draw=black!30,
                fill=white,
                /tikz/every even column/.append style={column sep=0.4cm},
            },
            legend image code/.code={%
                \draw[#1, draw=none] (0cm,-0.08cm) rectangle (0.3cm,0.16cm);
            },
            legend cell align={left},
            legend columns=1,
            nodes near coords,
            nodes near coords style={
                font=\tiny,
                /pgf/number format/fixed,
                /pgf/number format/precision=0,
                rotate=90,
                anchor=west,
                xshift=1pt,
                color=black!70,
            },
            ymajorgrids=true,
            grid style={dashed, gray!25},
            axis line style={gray!60},
            tick style={gray!60},
        ]
        \addplot[fill=barlow, draw=barlow!80!black] coordinates {
            (HunyuanVideo, 185.9)
            (HunyuanVideo-1.5, 212.9)
            (Wan2.1, 382.7)
            (Wan2.2, 383.0)
        };
        \addplot[fill=barmid, draw=barmid!80!black] coordinates {
            (HunyuanVideo, 411.4)
            (HunyuanVideo-1.5, 478.5)
            (Wan2.1, 855.4)
            (Wan2.2, 856.0)
        };
        \addplot[fill=barhigh, draw=barhigh!80!black] coordinates {
            (HunyuanVideo, 608.6)
            (HunyuanVideo-1.5, 705.9)
            (Wan2.1, 1269.6)
            (Wan2.2, 1274.5)
        };
        \legend{121 frames, 193 frames, 241 frames}
        \end{axis}
    \end{tikzpicture}
    \caption{Total energy consumption per model and frame count. All measurements use $8H2$ at 720$\times$1280 resolution, 40 denoising steps, and batch size 1.}
    \label{fig:ModelSizeEffectBatch}
\end{figure}

Testing multiple models sharing the same architecture but differing in parameter count allowed us to isolate the effect of model size on energy consumption, which we show in  Figure~\ref{fig:ModelSizeEffectBatch}. It reveals three important findings: (1) HunyuanVideo-1.5 (8.3 B) consistently consumes more energy than HunyuanVideo (13B) despite having a fewer number of parameters, 
whereas (2) Wan 2.1 (14B) and Wan 2.2 (27B [14B active via MoE]) are nearly indistinguishable at all configurations. 
(3) Despite having a comparable active parameter count, Wan 2.1 consumes, on average, roughly twice the energy of HunyuanVideo, underscoring that architectural design choices, not parameter count, determine computational cost. 

\begin{takeaway}
Total or active parameter count is not a consistent predictor of energy consumption or generation time. Consequently, estimating the model size of closed models is neither necessary nor sufficient to estimate the energy consumption.  
\end{takeaway}


Our analysis reveals that energy consumption per video remains nearly constant across all batch sizes (as seen in Appendix~\ref{appendix_batch_size}), confirming that \emph{energy consumption scales linearly with batch size} due to the predominantly compute-bound nature of diffusion models. For instance, generating a 720$H$, 1280$W$, 30$S$, and 121$F$ HunyuanVideo video consumes 146.1 $Wh$ at batch size 1, 141.3 $Wh$ at batch size 2, and 144.2 $Wh$ at batch size 4 with a maximum deviation of less than 3.3\%, with minor variations attributable to temporal tiling effects. Total generation energy thus scales linearly with batch size for video generation models, unlike for text generation models.

\begin{takeaway}
Total generation energy for T2V models scales linearly with batch size due to its compute-bound nature, with batching yielding no per-video efficiency gains. 

\end{takeaway}

\subsection{Architectural Decomposition \& Forward Validation}  \label{forward}

Based on the theoretical foundation and empirical observations above, we develop the formulas based on the architectural first principles of each model. Using these coefficients, we showcase the forward direction of our framework by fitting these formulas to the data, as well as validating the robustness of the framework on unseen configurations using cross-validation.

\subsubsection{
The Linear \& Quadratic Decomposition of Diffusion Models}  \label{subsec_decomposition}

Factoring in the theoretical foundation and empirical observations above, we are able to fit the energy scaling formulas using non-negative least squares (NNLS) regression, giving:$E_{\text{denoise}} = S \cdot B \cdot ( N_1 \cdot T^2 + M \cdot T ) + G \cdot B$ for the denoising stage; for the VAE stage, denoising steps do not affect decoding, so $E_{\text{VAE}} = B \cdot ( M \cdot T + G )$. However, since HunyuanVideo VAE employs a spatiotemporal mid-block attention mechanism (scales with $O(T^2)$) \cite{kong2024hunyuanvideo} with temporal tiling, the attention scaling turns to $(WHC)^2 \cdot F/C \propto W^2H^2F$-- meaning that it is quadratic in spatial resolution but linear in frame count~\footnote{$C$ denotes the fixed temporal tile size.}
-- resulting in $E_{\text{VAE}} = B \cdot ( N \cdot (W \cdot H)^2 \cdot F + M \cdot T + G )$. Since both stages share the same functional forms in $T$, and given the limited variation in $S$ across experiments, we fit a single unified formula on total energy, merging VAE contributions into the denoising coefficients to avoid collinearity, giving:

\begin{equation}
E = \underbrace{N_1 \cdot T^2 \cdot S \cdot B}_{\text{denoising self-attn}} + \underbrace{N_2 \cdot {(W \cdot H)^2 \cdot F} \cdot B}_{\text{VAE mid-block attn}} + \underbrace{M \cdot T \cdot S \cdot B}_{\text{FFN + VAE Conv}} + \underbrace{G \cdot B}_{\text{overhead}}
\end{equation}

where $N_1$ captures the denoising self-attention cost, $N_2$ the VAE spatiotemporal mid-block attention cost relevant only to HunyuanVideo (zero for all other models), $M$ both denoising FFN and VAE convolutions, and $G$ the fixed per-video overhead.

\begin{table}[h!]
\centering
\scriptsize
\resizebox{\textwidth}{!}{
\begin{tabular}{ccccccccccl}
\hline
\multicolumn{11}{c}{ $8H2$} \\ \hline
\textbf{Model} & \textbf{N1} & \textbf{SE N1} & \textbf{M} & \textbf{SE M} & \textbf{N2} & \textbf{SE N2} & \textbf{G} & \textbf{SE G} & \textbf{err \%} & \textbf{CV \%} \\ \hline
HunyuanVideo & 2.40E-10 & 1.82E-12 & 8.5E-06 & 5.80E-07 & 3.20E-13 & 1.63E-14 & 0.22 & 1.62 & 1.709 & 1.803 \\
HunyuanVideo-1.5 & 3.00E-10 & 1.10E-11 & 1.05E-05 & 3.20E-06 & -- & -- & 18.61 & 9.56 & 0.672 & 1.157 \\
Wan2.1 & 5.25E-10 & 2.42E-12 & 2.52E-05 & 6.90E-07 & -- & -- & 9.52 & 2.01 & 0.093 & 0.132 \\
Wan2.2 & 5.36E-10 & 1.01E-12 & 2.32E-05 & 4.01E-07 & -- & -- & 12.93 & 1.49 & 0.137 & 0.169 \\ \hline
\multicolumn{11}{c}{$H2$} \\ \hline
HunyuanVideo-1.5 & 2.82E-10 & 3.99E-12 & 8.29E-06 & 6.91E-07 & -- & -- & 0.00 & 0.00 & 0.154 & 0.471 \\
Wan2.1 & 5.42E-10 & 1.14E-12 & 1.64E-05 & 1.97E-07 & -- & -- & 2.29 & 0.30 & 0.021 & 0.085 \\
Wan2.2 & 5.48E-10 & 4.93E-12 & 1.67E-05 & 8.54E-07 & -- & -- & 1.85 & 1.32 & 0.091 & 0.365 \\
LTX-2-T2VA & 4.56E-13 & -- & 7.79E-07 & -- & 1.83E-11 & -- & 1.00 & -- & 2.065 & 2.088 \\
LTX-2-T2V& 4.56E-13 & 8.57E-14 & 6.47E-07 & 5.17E-08 & 1.83E-11 & 2.31E-12 & 0.56 & 0.44 & 1.896 & 2.431 \\ \hline
\multicolumn{11}{c}{$B2$} \\ \hline
HunyuanVideo-1.5 & 1.88E-10 & 2.87E-12 & 4.46E-06 & 4.97E-07 & -- & -- & 2.25 & 0.77 & 0.160 & 0.639 \\
Wan2.1 & 3.31E-10 & 1.38E-12 & 1.31E-05 & 2.38E-07 & -- & -- & 0.94 & 0.37 & 0.039 & 0.156 \\
Wan2.2 & 3.33E-10 & 2.33E-13 & 1.32E-05 & 4.03E-08 & -- & -- & 1.07 & 0.06 & 0.007 & 0.026 \\
LTX-2-T2VA & 2.29E-13 & -- & 6.11E-07 & -- & 1.35E-11 & -- & 1.01 & -- & 3.123 & 3.220 \\
LTX-2-T2V & 2.29E-13 & 6.26E-14 & 5.00E-07 & 3.77E-08 & 1.35E-11 & 1.68E-12 & 0.65 & 0.32 & 2.280 & 2.859 \\ \hline
\end{tabular}}
\caption{Estimated coefficients and accuracy across all measured models. err \% refers to MAPE and CV \% refers to cross-validation MAPE }
\label{tab: coeff}
\end{table}

Since LTX-2 model variants follow a two-stage generation pipeline,
the second-stage FFN term and VAE decoding share the same functional forms in $T$ as the first stage, so they are absorbed into the fitted coefficients to avoid collinearity. The second-stage self-attention cost at full resolution is instead captured by a separate $N_2 \cdot T^2 \cdot B$ term, reflecting the quadratic attention cost of the 3 full-resolution denoising steps independently of $S$. For LTX-2 models with no audio generation, we first fit the NS variant: 

\begin{equation}
E_{\text{NS}} =
\underbrace{N_1 \cdot T^2 \cdot S \cdot B}_{\text{1st-stage self-attn}}
+
\underbrace{N_2 \cdot T^2 \cdot B}_{\text{2nd-stage self-attn}}
+
\underbrace{M \cdot T \cdot S \cdot B}_{\text{FFN (1st \& 2nd stage)+VAE Conv}}
+
\underbrace{G \cdot B}_{\text{overhead}}
\end{equation}

The key difference between LTX-2-T2V and LTX-2-T2VA lies in the model's multimodal classifier-free guidance (CFG) formulation~\cite{ltx2}. For video-only generation, CFG requires two forward passes per step -- one conditioned on the text input and one on an empty or negative prompt -- to compute the guidance direction. Joint audio-video generation introduces a third pass to independently modulate cross-modal influence (with and without audio), bringing the total to 3 passes per step. Differencing T2V and T2VA models' measurements under identical configurations isolates this overhead as $\Delta M \cdot T \cdot S \cdot B + \Delta G \cdot B$. The audio waveform itself takes under 0.1 seconds to generate for 10 seconds of audio on a single H200 \cite{liao2024fish, voxtral_tts_2603} and is therefore not modeled; the overhead does not stem from the audio generation itself nor from the audio-video cross-attention term but entirely from the additional video transformer forward pass introduced by the multimodal CFG formulation. To account for this overhead in the full T2VA formula, we add  $\Delta M$ to the $M$ term and $\Delta G$ to the $G$ term.

\subsubsection{Forward Validation of the Scaling Laws}

\begin{figure}[t]
    \centering
    \definecolor{cA}{HTML}{E69F00} 
    \definecolor{cB}{HTML}{0072B2} 
    \definecolor{cC}{HTML}{D55E00} 
    \definecolor{cD}{HTML}{CC79A7} 
    \definecolor{cE}{HTML}{009E73} 
    \definecolor{cF}{HTML}{56B4E9} 
    \begin{tikzpicture}
        \begin{groupplot}[
            group style={
                group size=3 by 1,
                horizontal sep=1.0cm,
            },
            width=0.37\linewidth,
            height=5.2cm,
            xlabel={Actual energy (Wh)},
            ylabel={Predicted energy (Wh)},
            xlabel style={font=\footnotesize, yshift=2pt},
            ylabel style={font=\footnotesize, yshift=-2pt},
            tick label style={font=\scriptsize},
            title style={font=\small\bfseries, yshift=-1ex},
            grid=major,
            grid style={dashed, gray!20},
            axis line style={gray!60},
            tick style={gray!60},
            scaled ticks=false,
            tick align=outside,
            legend style={
                draw=black!25,
                font=\footnotesize,
                /tikz/every even column/.append style={column sep=0.45cm},
                fill=white,
                fill opacity=0.95,
                text opacity=1,
                rounded corners=1pt,
            },
            legend to name=fittedmodelslegend,
            legend columns=7,
        ]

        \nextgroupplot[
            title={$8H2$},
            xmin=0, xmax=3600,
            ymin=0, ymax=3600,
        ]
        \addplot[dashed, gray!60, thick, forget plot] coordinates {(0,0) (3600,3600)};
        \addplot[only marks, mark=*, color=cA, mark size=1.6pt, mark options={fill=cA, fill opacity=0.75, draw=black!55, line width=0.25pt}, forget plot] coordinates {
          (24.186,23.318)
          (47.915,46.499)
          (69.073,65.863)
          (131.767,127.964)
          (213.963,209.619)
          (30.226,29.799)
          (60.661,59.981)
          (88.006,85.317)
          (169.973,166.908)
          (277.778,274.572)
          (36.450,36.280)
          (73.605,73.463)
          (107.357,104.770)
          (208.297,205.852)
          (341.281,339.526)
          (146.107,151.109)
          (319.768,325.813)
          (185.909,190.434)
          (411.397,416.843)
          (608.576,614.714)
          (1260.930,1273.654)
          (2164.977,2167.253)
          (225.538,229.758)
          (502.941,507.873)
          (747.551,751.947)
          (1555.174,1567.460)
          (2677.995,2676.298)
          (46.939,46.636)
          (96.732,92.997)
          (136.686,131.727)
          (261.602,255.927)
          (425.500,419.237)
          (60.826,59.598)
          (122.964,119.962)
          (174.674,170.633)
          (337.721,333.816)
          (552.973,549.144)
          (73.423,72.559)
          (149.531,146.926)
          (212.863,209.540)
          (414.021,411.704)
          (680.247,679.051)
          (282.573,302.219)
          (632.013,651.626)
          (370.930,380.867)
          (816.882,833.686)
          (450.124,459.516)
          (999.040,1015.745)
          (91.518,93.273)
          (189.062,185.995)
          (267.022,263.453)
          (516.036,511.854)
          (844.083,838.474)
          (118.934,119.196)
          (240.784,239.923)
          (342.682,341.267)
          (666.926,667.631)
          (1096.374,1098.288)
          (143.638,145.118)
          (292.573,293.852)
          (418.312,419.080)
          (818.231,823.408)
          (1350.717,1358.103)
          (576.931,604.438)
          (1335.314,1303.252)
          (730.172,761.735)
          (1727.699,1667.371)
        };
        \addplot[only marks, mark=square*, color=cB, mark size=1.6pt, mark options={fill=cB, fill opacity=0.75, draw=black!55, line width=0.25pt}, forget plot] coordinates {
          (165.071,165.638)
          (212.913,214.646)
          (478.474,473.089)
          (705.931,704.160)
          (260.764,263.654)
          (589.638,586.708)
          (870.652,875.546)
        };
        \addplot[only marks, mark=pentagon*, color=cE, mark size=1.6pt, mark options={fill=cE, fill opacity=0.75, draw=black!55, line width=0.25pt}, forget plot] coordinates {
          (289.328,289.631)
          (382.691,383.001)
          (855.352,852.912)
          (1269.571,1268.849)
          (476.132,476.371)
          (1063.848,1063.760)
          (1582.124,1583.682)
          (289.196,289.631)
          (382.833,383.001)
          (476.135,476.371)
        };
        \addplot[only marks, mark=star, color=cF, mark size=1.6pt, mark options={fill=cF, fill opacity=0.75, draw=black!55, line width=0.25pt}, forget plot] coordinates {
          (290.158,290.335)
          (382.884,382.806)
          (855.987,855.915)
          (1274.504,1276.198)
          (2694.987,2693.973)
          (476.933,475.276)
          (1068.273,1066.664)
          (1589.161,1592.017)
          (3364.563,3364.236)
          (289.544,290.335)
          (382.955,382.806)
          (476.554,475.276)
          (289.241,290.335)
          (382.674,382.806)
          (475.835,475.276)
        };
        \node[draw=black!35, fill=white, fill opacity=0.92, text opacity=1, anchor=north west, font=\scriptsize, align=left, inner sep=3pt, rounded corners=1pt] at (rel axis cs:0.04,0.97) {MAPE = 1.23\%\\Max\phantom{PE} = 6.95\%};

        \nextgroupplot[
            title={$H2$},
            xmin=0, xmax=420,
            ymin=0, ymax=420,
            ylabel={},
        ]
        \addplot[dashed, gray!60, thick, forget plot] coordinates {(0,0) (420,420)};
        \addplot[only marks, mark=square*, color=cB, mark size=1.6pt, mark options={fill=cB, fill opacity=0.75, draw=black!55, line width=0.25pt}, forget plot] coordinates {
          (65.519,65.718)
          (149.304,149.168)
          (87.772,87.624)
          (198.787,198.890)
        };
        \addplot[only marks, mark=triangle*, color=cC, mark size=1.6pt, mark options={fill=cC, fill opacity=0.75, draw=black!55, line width=0.25pt}, forget plot] coordinates {
          (4.537,4.484)
          (7.162,7.047)
          (11.236,11.030)
          (14.422,14.443)
          (5.417,5.551)
          (8.620,8.815)
          (13.435,13.816)
          (17.497,18.053)
          (7.179,6.908)
          (11.772,11.612)
          (19.214,19.270)
          (25.677,26.051)
          (8.759,8.640)
          (14.413,14.541)
          (23.492,23.992)
          (31.417,32.267)
          (32.884,31.349)
          (39.889,38.689)
        };
        \addplot[only marks, mark=diamond*, color=cD, mark size=1.6pt, mark options={fill=cD, fill opacity=0.75, draw=black!55, line width=0.25pt}, forget plot] coordinates {
          (3.605,3.573)
          (5.869,5.835)
          (9.464,9.419)
          (12.409,12.532)
          (4.383,4.471)
          (7.271,7.335)
          (11.621,11.802)
          (15.039,15.639)
          (6.029,5.711)
          (10.071,9.947)
          (16.878,16.980)
          (22.750,23.293)
          (7.388,7.180)
          (12.433,12.456)
          (20.649,21.074)
          (27.890,28.723)
          (29.593,28.260)
          (58.069,58.421)
          (36.057,34.707)
          (70.541,70.651)
        };
        \addplot[only marks, mark=pentagon*, color=cE, mark size=1.6pt, mark options={fill=cE, fill opacity=0.75, draw=black!55, line width=0.25pt}, forget plot] coordinates {
          (129.541,129.590)
          (290.561,290.512)
          (172.059,172.023)
          (386.549,386.586)
        };
        \addplot[only marks, mark=star, color=cF, mark size=1.6pt, mark options={fill=cF, fill opacity=0.75, draw=black!55, line width=0.25pt}, forget plot] coordinates {
          (130.467,130.678)
          (293.616,293.405)
          (173.779,173.621)
          (390.433,390.591)
        };
        \node[draw=black!35, fill=white, fill opacity=0.92, text opacity=1, anchor=north west, font=\scriptsize, align=left, inner sep=3pt, rounded corners=1pt] at (rel axis cs:0.04,0.97) {MAPE = 1.52\%\\Max\phantom{PE} = 5.27\%};

        \nextgroupplot[
            title={$B2$},
            xmin=0, xmax=270,
            ymin=0, ymax=270,
            ylabel={},
        ]
        \addplot[dashed, gray!60, thick, forget plot] coordinates {(0,0) (270,270)};
        \addplot[only marks, mark=square*, color=cB, mark size=1.6pt, mark options={fill=cB, fill opacity=0.75, draw=black!55, line width=0.25pt}, forget plot] coordinates {
          (43.494,43.617)
          (97.867,97.744)
          (57.498,57.406)
          (129.484,129.576)
        };
        \addplot[only marks, mark=triangle*, color=cC, mark size=1.6pt, mark options={fill=cC, fill opacity=0.75, draw=black!55, line width=0.25pt}, forget plot] coordinates {
          (3.939,3.705)
          (5.931,5.598)
          (8.782,8.482)
          (11.191,10.917)
          (4.393,4.521)
          (6.731,6.932)
          (10.213,10.549)
          (13.079,13.563)
          (5.772,5.496)
          (9.058,8.900)
          (14.340,14.321)
          (18.532,19.048)
          (6.744,6.803)
          (10.716,11.067)
          (17.175,17.734)
          (22.401,23.470)
          (23.681,22.709)
          (44.123,44.319)
          (79.520,83.284)
          (29.074,27.878)
          (53.672,53.536)
          (96.028,99.036)
        };
        \addplot[only marks, mark=diamond*, color=cD, mark size=1.6pt, mark options={fill=cD, fill opacity=0.75, draw=black!55, line width=0.25pt}, forget plot] coordinates {
          (2.952,2.919)
          (4.658,4.559)
          (7.240,7.108)
          (9.360,9.291)
          (3.438,3.593)
          (5.502,5.668)
          (8.630,8.837)
          (11.315,11.515)
          (4.685,4.470)
          (7.686,7.481)
          (12.451,12.377)
          (16.403,16.710)
          (5.455,5.556)
          (9.145,9.295)
          (14.978,15.263)
          (19.698,20.474)
          (21.080,20.095)
          (40.012,40.370)
          (25.395,24.513)
          (48.399,48.390)
        };
        \addplot[only marks, mark=pentagon*, color=cE, mark size=1.6pt, mark options={fill=cE, fill opacity=0.75, draw=black!55, line width=0.25pt}, forget plot] coordinates {
          (85.551,85.610)
          (187.990,187.931)
          (113.876,113.832)
          (250.216,250.260)
        };
        \addplot[only marks, mark=star, color=cF, mark size=1.6pt, mark options={fill=cF, fill opacity=0.75, draw=black!55, line width=0.25pt}, forget plot] coordinates {
          (86.331,86.341)
          (189.438,189.428)
          (114.773,114.766)
          (252.208,252.215)
        };
        \node[draw=black!35, fill=white, fill opacity=0.92, text opacity=1, anchor=north west, font=\scriptsize, align=left, inner sep=3pt, rounded corners=1pt] at (rel axis cs:0.04,0.97) {MAPE = 2.13\%\\Max\phantom{PE} = 5.94\%};
        \addlegendimage{dashed, gray!60, thick}
        \addlegendentry{Perfect fit}
        \addlegendimage{only marks, mark=*, color=cA, mark size=1.8pt, mark options={fill=cA, fill opacity=0.75, draw=black!55, line width=0.25pt}}
        \addlegendentry{HunyuanVideo}
        \addlegendimage{only marks, mark=square*, color=cB, mark size=1.8pt, mark options={fill=cB, fill opacity=0.75, draw=black!55, line width=0.25pt}}
        \addlegendentry{HunyuanVideo-1.5}
        \addlegendimage{only marks, mark=triangle*, color=cC, mark size=1.8pt, mark options={fill=cC, fill opacity=0.75, draw=black!55, line width=0.25pt}}
        \addlegendentry{LTX-2-T2VA}
        \addlegendimage{only marks, mark=diamond*, color=cD, mark size=1.8pt, mark options={fill=cD, fill opacity=0.75, draw=black!55, line width=0.25pt}}
        \addlegendentry{LTX-2-T2V}
        \addlegendimage{only marks, mark=pentagon*, color=cE, mark size=1.8pt, mark options={fill=cE, fill opacity=0.75, draw=black!55, line width=0.25pt}}
        \addlegendentry{Wan2.1}
        \addlegendimage{only marks, mark=star, color=cF, mark size=1.8pt, mark options={fill=cF, fill opacity=0.75, draw=black!55, line width=0.25pt}}
        \addlegendentry{Wan2.2}

        \end{groupplot}
    \end{tikzpicture}
    \\[0.6em]
    \pgfplotslegendfromname{fittedmodelslegend}
    \caption{Predicted versus actual energy consumption (Wh) across all measured configurations. The dashed diagonal indicates the perfect-fit line. Per-panel MAPE and maximum APE shown inset; per-panel sample sizes are $n{=}99$, $n{=}50$, and $n{=}54$ from left to right.}
    \label{fig:fittedmodels}

\end{figure}
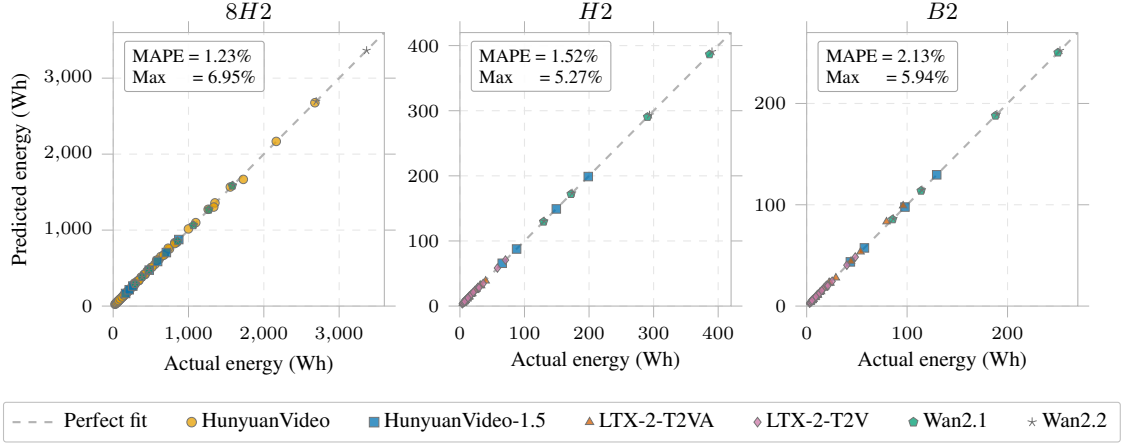

The fitted scaling laws achieve strong accuracy across all models and hardware configurations, as reported in Figure \ref{fig:fittedmodels} and Table~\ref{tab: coeff}. On the full $8H2$ station, Mean Average Percentage Error (MAPE) ranges from 0.093\% (Wan 2.1) to 1.709\% (HunyuanVideo), with maximum errors of 0.285\% and 6.953\%, respectively; the latter is driven by HunyuanVideo's VAE tiling effects at high resolutions and batch sizes, which introduce some variability due to the memory-bound nature of the decoding phase. On single GPUs, Wan models achieve a near-perfect fit (0.021--0.091\% MAPE on $H2$, 0.007-0.039\% on $B2$), while LTX-2 variants reach 1.896--3.123\%  MAPE and up to 5.943\% maximum due to the two-stage pipeline introducing additional variance across configurations. Additionally, the LTX-2 T2VA fitting accuracy showcases that the energy overhead of audio generation is driven mainly by the additional video transformer forward pass, not the audio terms themselves. Across all models and hardware configurations, the leave-one-out cross-validation MAPE remains within a small margin of the in-sample MAPE (e.g., 1.803\% vs.\ 1.709\% for HunyuanVideo, 0.132\% vs.\ 0.093\% for Wan2.1), confirming that the fitted scaling coefficients are stable and generalize reliably to unseen configurations, making this framework a robust basis to compare models.

In terms of full node testing, although the $8H2$ configuration generates videos at substantially shorter times (roughly $8\times$ faster than $H2$), the compute-bound nature of the process means that doubling the number of GPUs roughly halves generation time while simultaneously doubling power draw, leaving total energy consumption similar or higher due to inter-GPU communication overhead. This overhead manifests directly in the fitted $M$ coefficient, which is higher on the full station than on a single GPU (2.52$\times 10^{-5}$ vs. 1.64$\times 10^{-5}$ for Wan 2.1; 2.32$\times 10^{-5}$ vs. 1.67$\times 10^{-5}$ for Wan 2.2), consistent with the inter-GPU synchronization cost growing with the number of tokens processed. At 40 steps and 121 frames, this translates to 383 vs. 345 $Wh$ for Wan 2.1, 383 vs. 349 $Wh$ for Wan 2.2, and 215 vs. 177 $Wh$ for HunyuanVideo-1.5, revealing a consistent energy penalty for multi-GPU inference.

\begin{takeaway}
The energy consumption profile of video diffusion models is fully characterized by the quadratic scaling of self-attention, the linear scaling of FFN, VAE convolutions, batch size, and denoising steps. This confirms that energy consumption and thus generation time are determined by architectural decisions and observable generation parameters alone.

\end{takeaway}
 \begin{takeaway}
 Multi-GPU inference is a latency optimization solution, not an energy one. Additionally, the energy overhead of joint audio-video generation is driven by the additional video forward pass required by the  CFG, not the audio terms themselves.
\end{takeaway}

\subsection{Backward Validation via Architectural Ablation}
\label{backward}

The backward direction of the framework rests on a single principle: fit quality is diagnostic of architectural correctness. A scaling law derived from the wrong architectural assumptions will therefore fail to capture the computational structure of the forward pass, manifesting as degraded accuracy. We demonstrate this by ablating architectural terms for two models -- HunyuanVideo on the full $8H2$ station and LTX-2-T2V on a single $H2$ -- and observing how fit quality responds.

For HunyuanVideo, removing the VAE mid-block attention term ($N_2$) causes a moderate degradation from 1.709\% to 3.768\% mean MAPE (6.953\% to 10.426\% max), reflecting the secondary but real contribution of the spatiotemporal tiled attention in the decoding phase. Further removal of the FFN term ($M$) while keeping the VAE term absent causes a collapse to 12.50\% mean and 81.32\% max MAPE. The same pattern holds for LTX-2-T2V: removing the second-stage attention term ($N_2$) degrades fit from 1.896\% to 4.424\% mean MAPE (5.267\% to 14.987\% max), and additionally, removing the primary FFN term collapses accuracy to 14.215\% mean and 80.012\% maximum MAPE. In both cases, this collapse indicates a systematic failure: the formula can no longer capture how energy scales, because an imperative term governing that scaling has been removed.

\begin{takeaway}
A scaling law only achieves strong accuracy when it correctly reflects the true model architecture. This indicates that the fit quality is its validation criterion -- i.e., the formula that best fits inference times is the one whose architectural assumptions are correct. 
\end{takeaway}


\subsection{Case Study: Extending the Framework to Proprietary Models}
\label{subsec_prop}

While direct energy measurement is infeasible for proprietary T2V systems such as Sora~\cite{sora2024} and Veo~\cite{veo2025}, the framework established above forms a complete chain that enables this estimation:  since video diffusion models run near TDP making time and energy interchangeable (Property 1), energy is governed by architectural scaling behavior rather than parameter count (Property 2); API providers are assumed to operate at batch size 1 (Property 3), and energy is fully characterized by generation parameters (Property 4). In principle, Properties 1 and 3 alone already reduce closed-model energy estimation to $E \approx P \cdot t$ simply multiplying observed latency by a near-TDP power value. However, we go further; by fitting the full scaling law formula to observed inference times, we aim to recover the architectural scaling behavior of each model, enabling energy prediction across arbitrary unseen configurations based on predictive accuracy reflecting architectural validity (Property 6). We pursue this as a case study to demonstrate the framework's broader applicability, while flagging upfront that the resulting estimates carry additional uncertainty from assumed hardware deployment and power draw -- neither of which can be directly verified for closed APIs. We evaluate 8 API-based T2V models, combining publicly available architectural information with generation-time measurements collected via the Fal.ai API~\cite{falai2026}. Specific per-model formulas, hardware deployment assumptions, and the Monte Carlo procedure used to propagate uncertainty are detailed in Appendix~\ref{prop_model}.

Our estimations show that the energy cost of generating a single 8-second 720p video can span over an order of magnitude between models: Veo 3 consumes 19.8--43.4 $Wh$ (mean 30.8 $Wh$), Seedance-1 72.1--87.9 $Wh$ (mean 80.0 $Wh$), Gen-4.5 241.6--410.9 $Wh$ (mean 322.0 $Wh$), and Sora 2.0 Pro 315.1--534.4 $Wh$ (mean 418.5 $Wh$). Veo 3's efficiency is consistent with its reported TPU v6e deployment (mean machine power 2.173 $kW$ \cite{tpu_paper}, well below DGX-class GPUs at 5.6--10 $kW$), while Seedance's lead among GPU-based models (2.5$\times$ less than Gen 4.5) suggests architectural choices matter as much as hardware. Gen 4.5 is the most methodologically clean estimate: Runway explicitly separates queuing time from generation time in API responses, yielding a near-perfect time fit (MAPE $<$0.10\%) that directly validates the backward estimation procedure on a single proprietary model where ground-truth-adjacent timing is available.
Sora 2.0 Pro is the most energy-intensive model in our case study, requiring on average 1{,}313~Wh per 12-second 1080p video. This is consistent with the fact that OpenAI announced the discontinuation of Sora in March 2026~\cite{futurism2026sora}, and the energy profile estimated here offers a view on its costs: 4M users~\cite{shrivastava_sora_2025} generating two 720p 8-second videos per day over six months would require an estimated 453.8--769.5~GWh (mean 602.6~GWh) -- equivalent to the consumption of 84k--142k US households~\cite{eia_faq_2024}. Across the 8 commercial systems evaluated, this case study illustrates that the framework can be extended to API-only models when hardware and power-draw uncertainty can be estimated via Monte Carlo approaches.

\section{Discussion and Limitations} \label{sec_discussion}
This study presents a framework to empirically characterize the energy required by T2V models across a range of architectures, obtained via prompting and direct energy measurement and extended  to proprietary systems via API-derived generation times. This contributes to a more rigorous empirical and architectural understanding of a modality whose footprint continues to grow. While video generation may represent a small fraction of overall AI usage today, adoption is accelerating -- driven by integration into established platforms such as Meta~\cite{metavibes2025} and OpenAI~\cite{sora2025}, as well as the emergence of platforms dedicated to AI-generated content like Pika~\cite{goldman2025}. Furthermore, AI is increasingly used in the advertising domain -- as illustrated by the AI-generated Coca-Cola ad, whose final product reportedly entailed prompting and refining over 70,000 videos~\cite{weatherbed2025}.

With that being said, we acknowledge that our study has two principal limitations. Firstly, the memory-bound nature of the decoding phase is not fully captured by our derived scaling laws, a direction that we are currently exploring for more robust estimations during the decoding phase. Secondly, due to high collinearity between terms, VAE and FFN cost cannot be entangled clearly, hindering phase-level coefficient interpretation; expanding the benchmark to a broader set of configurations by varying the generation parameters more independently would reduce collinearity among design matrix terms, yielding better-identified coefficients and tighter uncertainty bounds. Finally, future research is needed to map the factors that we have established to more types of models, modalities, and use cases (e.g., upscaling, object removal, image-to-video), as well as to further explore the various factors that influence proprietary models. We are actively pursuing these research directions and hope to publish further results and insights on the topic in the near future to build a ground basis for sustainability reporting in the AI community through empirically-grounded frameworks.

\clearpage
\bibliographystyle{unsrt}
\bibliography{biblio}

\clearpage
\appendix

\section*{Supplementary Materials}

\section{Packages \& Libraries}

\begin{itemize}
    \item \texttt{Python == 3.12}
    \item \texttt{torch == 2.10.0}
    \item \texttt{torchao == 0.16.0}
    \item \texttt{xDiT == 0.4.5} (Cloned from GitHub)
    \item \texttt{diffusers == 0.37.0.dev0} (Cloned from GitHub)
    \item \texttt{transformers == 4.57.6}
    \item \texttt{accelerate == 1.12}
    \item \texttt{pynvml == 13.0.1}
    \item \texttt{huggingface hub == 0.36.2}
    \item \texttt{OpenCV == 4.13.0}
    \item \texttt{numpy == 2.4.2}
    \item \texttt{pandas == 3.0.0}

\end{itemize}

\section{Hardware Details} \label{hardware}

\begin{itemize}
    \item \textbf{8H2}: a DGX H200 node equipped with 8 Nvidia H200 SXM GPU (5.6 $kW$ TDP, 1128GB VRAM, 3016GB RAM, 192vCPUs)
    \item \textbf{H2}: one H200 SXM GPU (0.7 $kW$ TDP, 141GB VRAM, 181GB RAM, 12vCPUs)
    \item \textbf{B2}: one B200 GPU (1 $kW$ TDP, 180GB VRAM, 283GB RAM, 28vCPUs)
\end{itemize}

\section{List of Prompts} \label{appendix_prompts}

\begin{itemize}
    \item In a symmetrical Wes Anderson style, a vintage library where all the books suddenly begin to float and rotate in mid-air. A young librarian with glasses looks up in awe. Static camera, pastel color palette, soft natural lighting through tall windows, perfectly centered framing.
    \item A cinematic, slow-motion tracking shot of a crystal glass falling onto a marble floor. The glass shatters into hundreds of sharp, reflective shards that fly toward the camera, while red wine splashes in realistic fluid ribbons across the white marble. High-speed photography style, 4K, realistic physics.
    \item A high-speed FPV drone shot weaving through a dense cyberpunk market at night. The camera zips between neon-lit stalls, under holographic signs, and through clouds of thick white steam rising from street food vendors. Motion blur, vibrant blue and magenta lighting, wet pavement reflections.
    \item     A group of penguins on a bright Antarctic ice shelf performing a synchronized rhythmic dance. They are flapping their flippers, hopping in place, and kicking up powdery snow. Cinematic lighting, 4k, high detail, realistic textures, soft sunlight reflecting off the ice.
    \item  Exactly two astronauts in white EVA suits walking on the surface of Mars during a massive dust storm. Red sand whips around their boots. One astronaut stops to pick up a glowing blue mineral. The camera does a slow 360-degree orbital rotation around them, maintaining sharp focus on the mineral.

\end{itemize}
\clearpage

\section{Batch Size Effect}
\label{appendix_batch_size}

\begin{figure}[h!]
    \centering
    \includegraphics[width=1\linewidth]{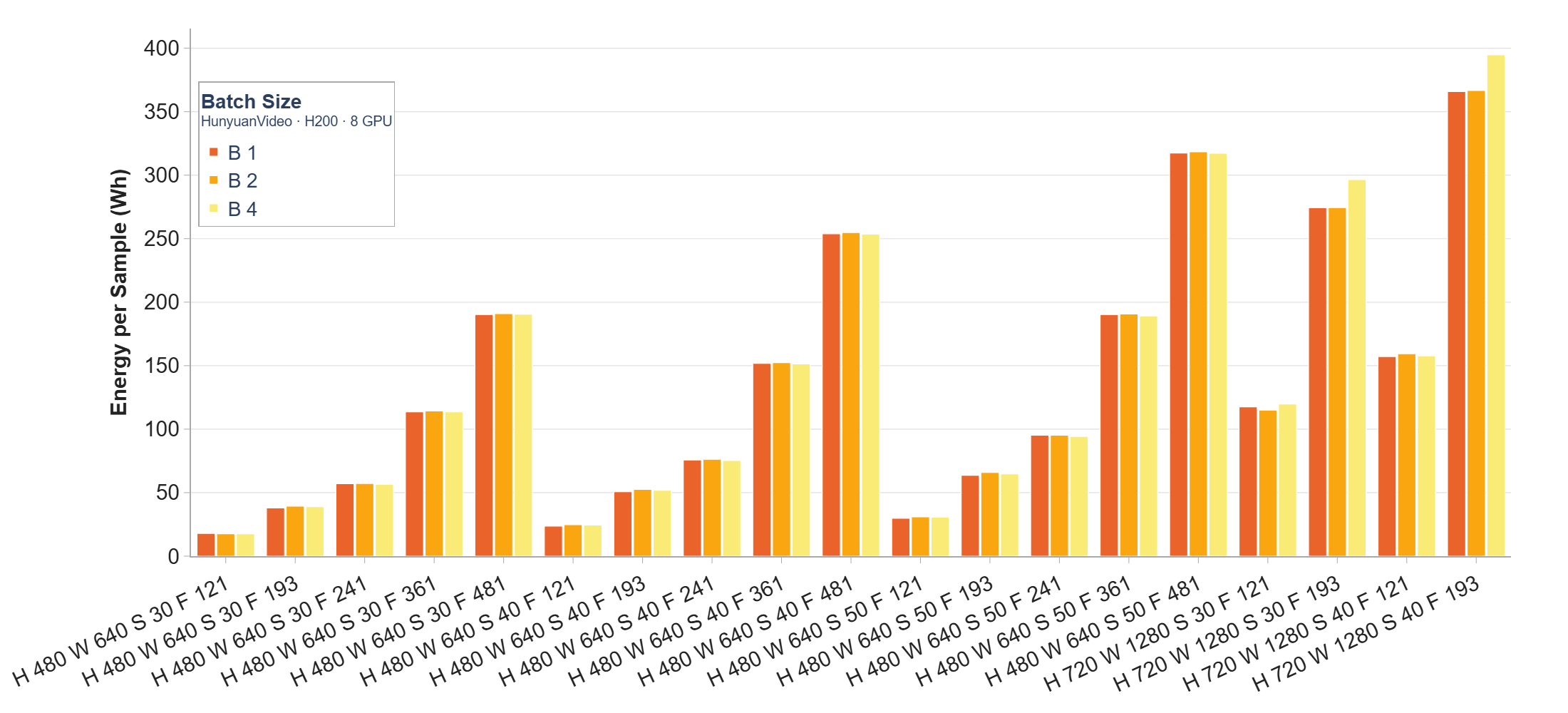}
    \caption{Energy Consumption for HunyuanVideo per configuration normalized by batch size (Energy Consumption / Batch Size)}
    \label{fig:BatchSize}
\end{figure}

Figure~\ref{fig:BatchSize} showcases the energy consumption per video for HunyuanVideo across different batch sizes on $8H2$. These results showcase that energy consumption scales linearly with batch size with no per-video efficiency gains.

\section{Proprietary Model Estimation: Methodology and Per-Model Derivations}
\label{prop_model}
This appendix details the methodology used in Section~\ref{subsec_prop} to estimate energy consumption for proprietary T2V models. We first describe the measurement procedure, hardware deployment assumptions, and Monte Carlo uncertainty propagation, then present the per-model formula derivations and full results.

\subsection{Methodology}

\paragraph{Measurement procedure.} For each of the 8 proprietary models, we query the API across 3--9 configurations spanning resolution and frame count (Table~\ref{tab: prop_table}). Because API latency conflates true generation time with queuing delays, each configuration was measured three times and the minimum latency retained -- queuing effects are strictly additive, so the minimum most faithfully approximates true generation time. We did not run identical configurations sequentially: the full configuration set was run, then re-run, to ensure variability in queueing conditions for any given configuration. Coefficients are fit to these minimum times using the same NNLS procedure applied to open models in Section~\ref{forward}.

\paragraph{Denoising steps.} For all proprietary models, we assume the number of denoising steps is fixed across configurations (consistent with API-level consistency guarantees). Since energy scales linearly with $S$ by construction, step count is absorbed into the fitted coefficients and treated as a model-level constant.

\paragraph{Hardware deployment assumptions.} We assume all models are deployed on DGX or equivalent multi-accelerator infrastructure to support user-facing generation times. This is consistent with our open-model benchmarks: on a full $8H2$ station, generating a 5-second 720p video already takes 105~s for HunyuanVideo and 190~s for Wan 2.1/2.2 -- single-GPU generation would be proportionally slower and incompatible with consumer expectations. The consistently long queuing times observed across proprietary APIs further support this assumption, as queuing overhead is characteristic of high-demand shared infrastructure rather than single-device deployment. Specific assignments:
\begin{itemize}
    \item \textbf{Veo models}: TPU v6e node (8 TPUs), reflecting Google's latest production infrastructure~\cite{tpu_paper}.
    \item \textbf{Seedance 1 and 1.5}: DGX H800 ($8H8$), reflecting GPU export restrictions limiting access to newer Nvidia hardware in China~\cite{reuters2026nvidiah200}.
    \item \textbf{All remaining models}: DGX B200 ($8B2$) or DGX H200 ($8H2$), with equal deployment likelihood assigned to each.
\end{itemize}

\paragraph{Power-draw model and Monte Carlo uncertainty.} Energy is estimated by modeling power draw as $P \sim \mathcal{N}(0.9 \cdot \text{TDP},\ (0.05 \cdot \text{TDP})^2)$, consistent with Property 1 and accounting for possible tiling effects. Where multiple hardware configurations are plausible candidates, each is assigned a deployment likelihood weight and sampled independently. Running 10{,}000 Monte Carlo simulations per model yields a weighted energy distribution and 95\% confidence interval for each configuration.

\paragraph{Audio-video models.} Property 4 implies no dedicated audio terms need to be estimated -- the overhead is absorbed into the fitted linear coefficient. For models with native audio generation, all measurements were collected with audio enabled. To estimate video-only energy, we use the proportional pricing between audio-video and video-only API tiers as a proxy; while absolute prices likely reflect competitive pressures rather than true cost, we assume the ratio is indicative of the relative computational overhead. We note explicitly that this is a proxy and less scientifically grounded than the audio-video estimations.

\subsection{Per-Model Formula Derivations}

\paragraph{Veo variants.} Veo models employ a decoupled spatial-temporal attention mechanism rather than standard joint spatiotemporal self-attention~\cite{veo3}. Instead of scaling as $T^2 \propto W^2H^2F^2$, the attention decomposes into a spatial term scaling as $WH^2 \cdot F$ and a temporal term scaling as $F^2 \cdot WH$. Given the limited variability in frame counts across tested configurations, the temporal term is collinear with the spatial term and cannot be reliably disentangled; we therefore retain only the dominant spatial attention term, yielding $E = N \cdot F \cdot (WH)^2 + M \cdot T + G$. For Veo 3.1, inference time measurements were heavily contaminated by queuing noise despite the min-of-3 procedure, making the attention terms unreliable; we therefore fit a linear formula $E = M \cdot T + G$ only, and note that this variant carries lower estimation accuracy than the others.

\paragraph{Seedance 1.} Seedance 1 follows a two-stage pipeline: primary generation at 480p followed by upscaling to the target resolution with fewer denoising steps~\cite{seedance1}. The full formula would be $E = N_1 \cdot F^2 + M_1 \cdot F + M_2 \cdot T + N_2 \cdot T^2 + G$, where $N_1$ captures the 480p self-attention cost and $M_1$ the corresponding FFN. Given the limited configuration variability and the likely negligible contribution of the second-stage attention, both the $N_2$ and $M_1$ terms are dropped to avoid collinearity, giving $E = N_1 \cdot F^2 + M \cdot T + G$.

\paragraph{Seedance 1.5.} Seedance 1.5 also employs decoupled attention~\cite{seedance1.5}, yielding a full formula of $E = M \cdot T + N_1 \cdot WH \cdot F^2 + N_2 \cdot F \cdot WH^2 + G$. To avoid collinearity between the two attention terms, we decouple them by retaining $N_1 \cdot F^2$ and $N_2 \cdot WH^2$ independently, dropping the shared $WH$ and $F$ prefactors that cannot be separately identified from the available configurations, giving $E = N_1 \cdot F^2 + M_2 \cdot WH^2 + G$.

\paragraph{Sora 2.0 Pro.} Sora 2.0 Pro natively supports up to 1080p but is known to generate at a lower internal resolution before upscaling -- the API explicitly distinguishes between 1080p and ``true'' 1080p~\cite{sora2}. We model the generation cost as a combination of a fixed lower-resolution term scaling linearly with $F$ and a full-resolution attention term with its corresponding FFN, dropping the lower-resolution attention due to its negligible and collinear contribution, giving $E = M_1 \cdot F + N_2 \cdot T^2 + M_2 \cdot T$.

\paragraph{Gen 4.5.} Gen 4.5 uses a standard spatiotemporal self-attention architecture with no documented multi-stage or decoupled design, and is fit with the standard formula $E = N \cdot T^2 + M \cdot T$. This is supported by Property 5: the accuracy of the formula reflects the validity of the architectural assumptions.

\subsection{Full Results}

\begin{table}[H]
\centering
\begin{tabular}{cclcclcc}
\hline
\textbf{Model} & \textbf{API} & \textbf{Capabilities} & \textbf{FPS} & \textbf{Count} & \textbf{Hardware} & \textbf{err \%} & \textbf{err \% max} \\ \hline
Veo 3.1        & Fal.AI       & V/A                   & 24           & 6              & TPUv6e            & 9.65\%          & 18.21\%          \\
Veo 3.1 Fast   & Fal.AI       & V/A                   & 24           & 6              & TPUv6e            & 11.09\%         & 23.17\%          \\
Veo 3          & Fal.AI       & V/A                   & 24           & 6              & TPUv6e            & 8.22\%          & 20.80\%          \\
Veo 3 Fast     & Fal.AI       & V/A                   & 24           & 6              & TPUv6e            & 10.33\%         & 19.81\%          \\
Sora 2.0 Pro      & Fal.AI       & V/A                   & 24           & 6              & $8B2$, $8H2$      & 3.99\%          & 9.45\%           \\
Seedance 1.0   & Fal.AI       & V                     & 24           & 9              & $8H8$             & 3.08\%          & 8.57\%           \\
Seedance 1.5   & Fal.AI       & V/A                   & 24           & 5              & $8H8$             & 3.80\%          & 7.28\%           \\
Gen 4.5        & Runway       & V                     & 24           & 3              & $8B2$, $8H2$      & 0.08\%          & 0.10\%           \\ \hline
\end{tabular}
\caption{List of proprietary models tested. \textbf{V}: Video only; \textbf{V/A}: Video and Audio. Hardware abbreviations: $8B2$ (DGX B200), $8H2$ (DGX H200), $8H8$ (DGX H800).}
\label{tab: prop_table}
\end{table}

\begin{figure}[H]
    \centering
    \includegraphics[width=1\linewidth]{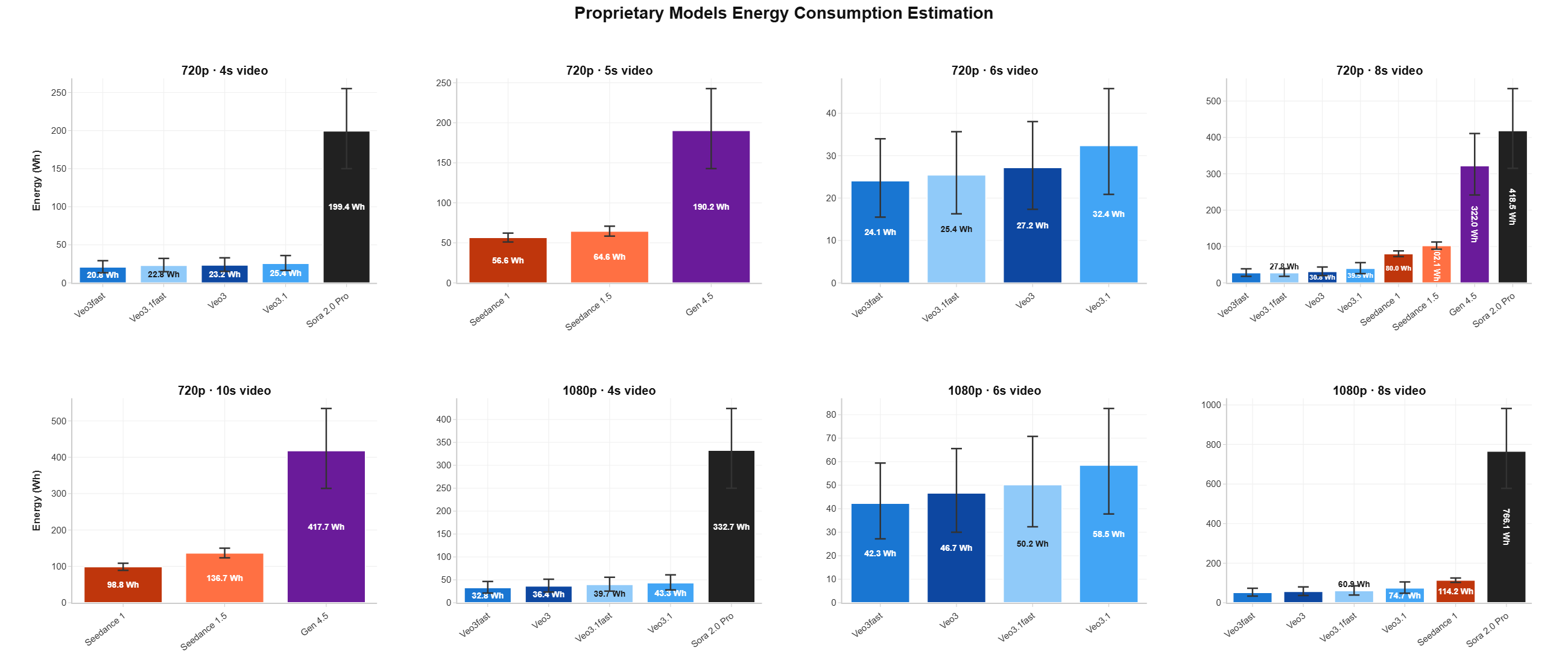}
    \caption{Estimated energy consumption (Wh) for all proprietary models across configurations}
    \label{fig:prop_plot}
\end{figure}

\end{document}